\documentclass[11pt,a4paper]{article}
\usepackage{moriond,epsfig,amsmath,floatflt}

\def\Journal#1#2#3#4{{#1} {\bf #2}, #3 (#4)}

\def\NPB{{\em Nucl. Phys.} B}
\def\PLB{{\em Phys.\@ Lett.\@ }  B}

\def\PRD{{\em Phys. Rev.} D}

\def\EPC{{\em Eur.\@ Phys.\@ J.\@ }C}
\def\CPC{{\em Comput.\@ Phys.\@ Commun.\@ }}
\def\ea{{\em et al.}}

\def\be{\begin{equation}}
\def\ee{\end{equation}}
\def\bea{\begin{eqnarray}}
\def\eea{\end{eqnarray}}

\begin{document}

%\initfloatingfigs

\vspace*{4cm}
\title{
\mathversion{bold}
JET PRODUCTION IN $ep$ COLLISIONS
\mathversion{normal}}

\author{PIERRE VAN MECHELEN \footnote{Postdoctoral fellow of the Fund
for Scientific Research - Flanders (Belgium);
Pierre.VanMechelen@ua.ac.be}}

\address{Universiteit Antwerpen}

\maketitle\abstracts{Results obtained by the H1 and ZEUS Collaborations
on jet production in photoproduction and deep-inelastic scattering are
presented and compared to NLO QCD models.}

\section{Introduction}

The observation of jet production at HERA allows to test the theory of
strong interactions, quantum chromodynamics (QCD), and to extract
information on the parton content of the proton and of real or virtual
photons. QCD predicts the production of partons with large transverse
momenta, fragmenting into jets with similar four-momenta.  The study
of jet observables therefore allows to investigate the underlying
parton dynamics. Perturbative QCD predictions, however, also need the
parton content of the proton and photon as input. Jet cross sections
can thus be used to further constrain the parton density functions
(PDF's) obtained by global fits.

The clustering of final state objects into a few jets is performed by
applying a jet finding algorithm. The results presented in this paper
are all obtained with the so-called inclusive, longitudinal invariant
$k_\perp$ algorithm~\cite{bib:ellis}.  This algorithm has the advantage
of being infrared and collinear safe and to be minimally sensitive to
fragmentation and underlying event effects. 

Reconstructed data are corrected from detector to hadron level using a
full detector simulation. Next-to-leading order (NLO) QCD predictions
are corrected from parton to hadron level.  This involves the
fragmentation of partons into hadrons and secondary interactions
between partons of the photon and proton remnants.  Correction factors
are obtained from leading order (LO) Monte Carlo (MC) models where NLO
effects are modelled by QCD cascades.

In LO two type of processes are distinguished.  In direct processes the
exchanged photon interacts as a whole with the proton to produce jets. 
In resolved interactions, the photon is treated as a source of partons,
one of which produces a hard scattering with the proton, leaving a
(soft) photon remnant behind.  The concept of resolved photons is
useful in photoproduction as well as in deep-inelastic scattering (DIS)
when $E_T^2 \gg Q^2$, where $E_T$ is the transverse energy of the
partons produced in the hard interaction and $Q^2$ is the virtuality of
the exchanged photon.  To separate direct and resolved enhanced event
samples, the momentum fraction of the photon entering the hard
scattering, $x_\gamma$, can be used.  On hadron level, the variable
$x_\gamma^{jet}$, which is correlated to the parton level $x_\gamma$,
is calculated as $x_\gamma^{jet} = {\sum_{jets} E_T^{jet} {\rm
e}^{-\eta^{jet}}} / {2 E_\gamma}$ (with $E_T^{jet}$ and $\eta^{jet}$
the jet transverse energy and pseudorapidity and $E_\gamma$ the
exchanged photon energy).

Using QCD factorization, the direct and resolved cross sections for
producing $N$ jets, integrated over phase space, can be expressed as

\begin{align}
\sigma_{direct}^{ep \rightarrow e + N jets + X} &= \int_\Omega d\Omega
f_{\gamma/e} (y, Q^2) \sum_i f_{i/p} (x_p, \mu_p^2) \sigma^{\gamma
i \rightarrow N jets}, \\
\sigma_{resolved}^{ep \rightarrow e + N jets + X} &= \int_\Omega d\Omega
f_{\gamma/e} (y, Q^2) \sum_{i j} f_{i/p} (x_p, \mu_p^2) f_{j/\gamma}
(x_\gamma, \mu_\gamma^2) \sigma^{i j \rightarrow N jets}.
\end{align}

\noindent Here $f_{\gamma/e}$, $f_{i/p}$ and $f_{j/\gamma}$ are flux
factors for photons originating from the electron and for partons
originating from the proton and the photon, respectively, evaluated at
given fractional momentum-energies and factorization scales.  The
partonic cross sections $\sigma^{\gamma i \rightarrow N jets}$
and $\sigma^{i j \rightarrow N jets}$ can be calculated in LO or
NLO as a function of the strong coupling constant $\alpha_S(\mu_R)$,
where $\mu_R$ is the renormalization scale.  The choice of
renormalization and factorization scales will lead to some uncertainty
in the predicted cross sections.  Different NLO calculations further
differ mainly in their treatment of infrared and collinear divergences.

\section{Inclusive jet photoproduction}

\begin{figure}[!htb]
\begin{center}
\begin{picture}(360,245)
\put(0,0){\psfig{figure=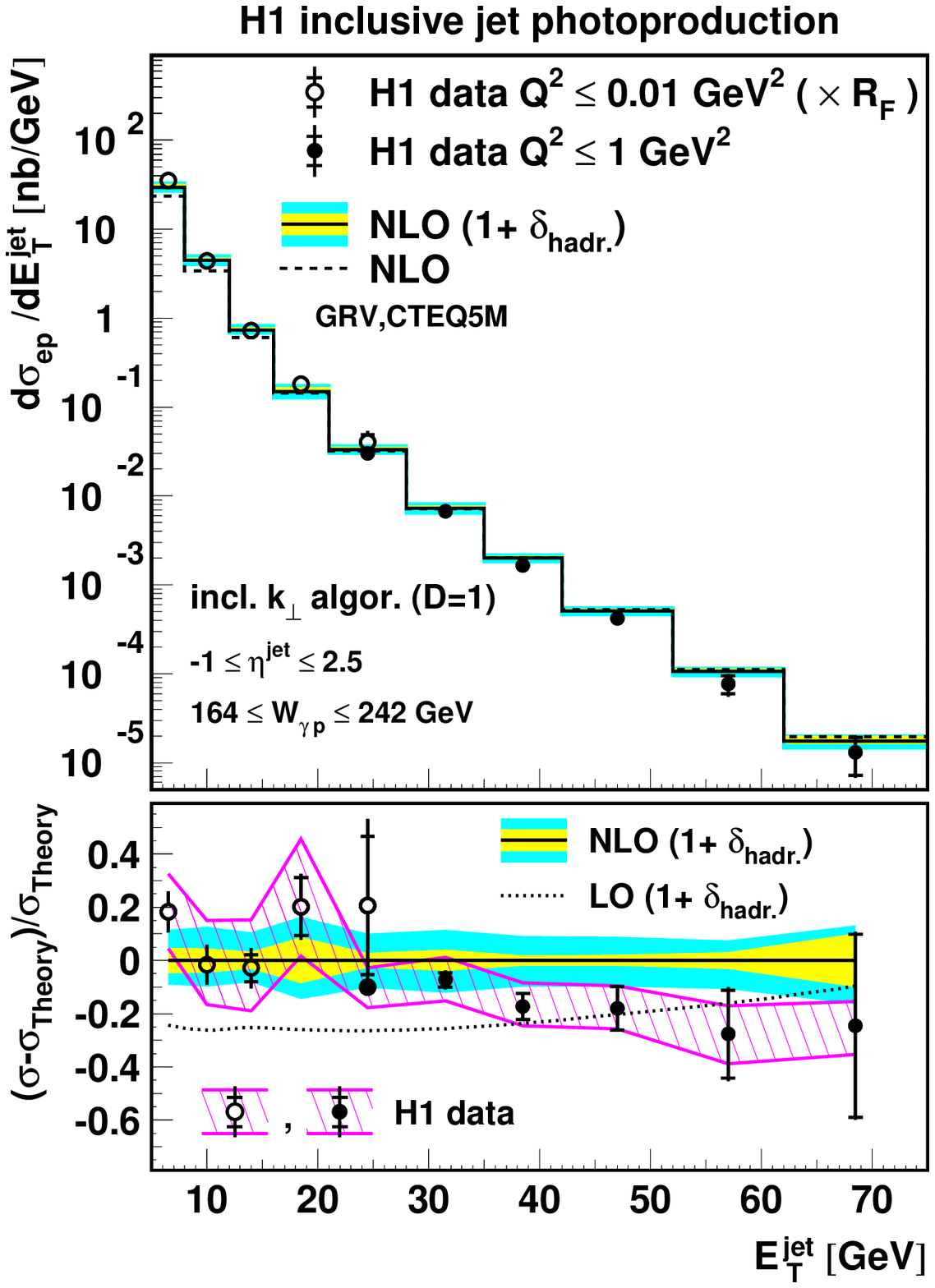,width=0.394\textwidth}}
\put(160,27){(a)}
\put(180,0){\psfig{figure=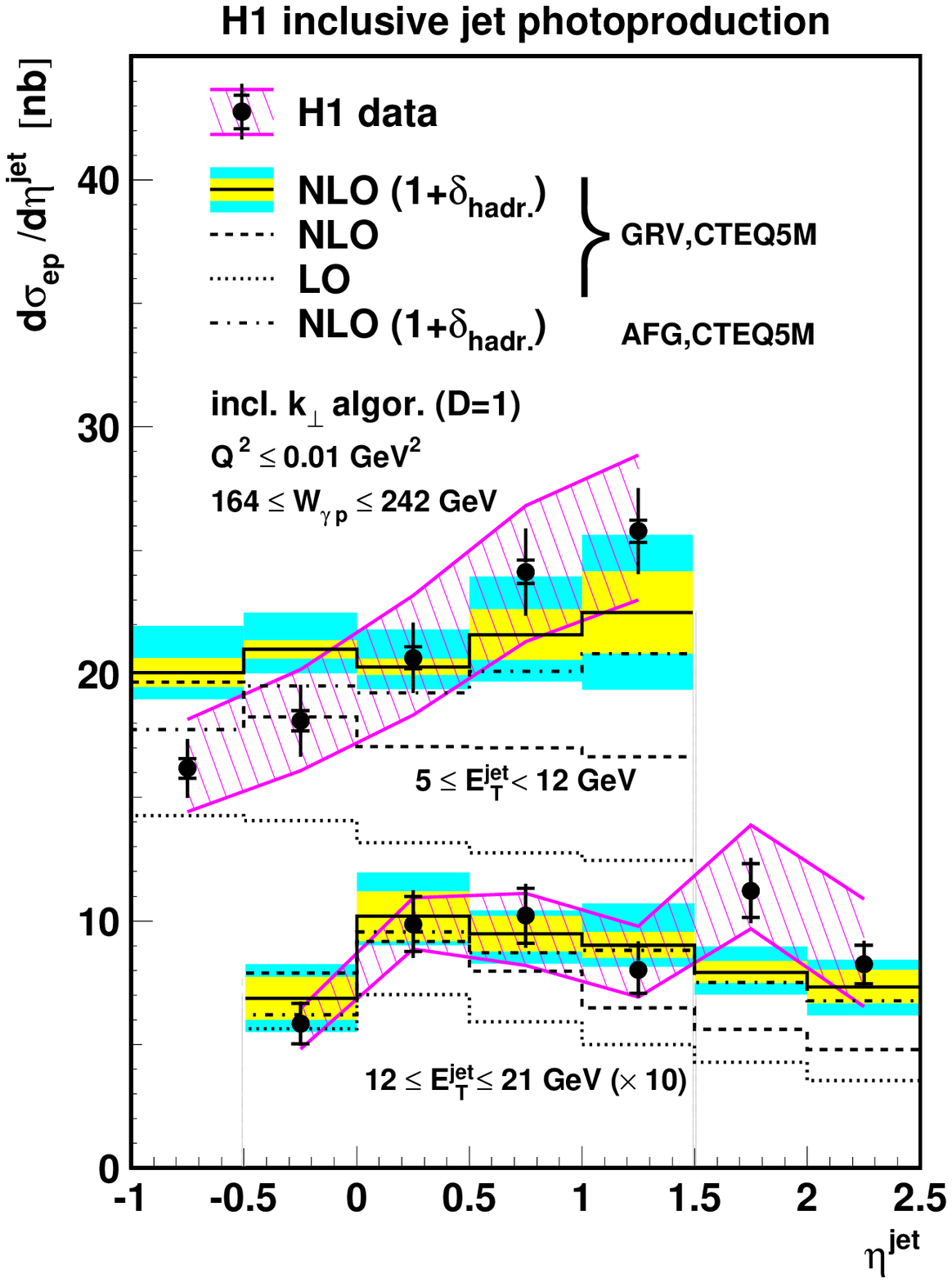,width=0.4\textwidth}}
\put(342,27){(b)}
\end{picture}
\end{center}

\caption{Differential inclusive jet cross sections as a function of
$E_T^{jet}$ (a) and $\eta^{jet}$ (b).  The hatched band shows the
uncertainty associated with the LAr calorimeter energy scale.  The
inside and outside shaded bands represent the uncertainty from
hadronisation corrections and from renormalization and factorization
scales, respectively.}

\label{fig:gammap_jets}

\end{figure}

Figure~\ref{fig:gammap_jets} shows the first H1 data on inclusive jet
photoproduction obtained with the $k_\perp$
algorithm~\cite{bib:h1gammapjets}.  This analysis comprises two data
sets. At large $E_T^{jet}$, tracker and calorimeter information is
emploid  to trigger the data acquisition, while at low $E_T^{jet}$ a
trigger based on the presence of the scattered electron in an electron
tagger is used, since in this kinematic range the first trigger suffers
from large backgrounds.  The low $E_T^{jet}$ data points are corrected
for the different $Q^2$ range covered, whereafter they agree well with
the high $E_T^{jet}$ data in the overlap region. 

The cross section as function of $E_T^{jet}$ is displayed in
Fig.~\ref{fig:gammap_jets}a.  It falls over more than six orders of
magnitude from $E_T^{jet} = 5$ to 75 GeV.  LO QCD predictions
underestimate the cross section (even with hadronisation corrections),
especially at low $E_T^{jet}$, while NLO QCD models reproduce the data well
after applying hadronisation corrections.  Different choices of photon
and proton PDF's vary at the level of 5 to 10 \% and describe the data
within errors.

Figure~\ref{fig:gammap_jets}b shows the cross section as a function of
$\eta^{jet}$, in two ranges of $E_T^{jet}$.  It is observed that
hadronisation corrections increase towards the proton remnant
direction. For $E_T^{jet} > 12 {\rm\ GeV}$, the data are well described
by NLO prediction, but at lower $E_T^{jet}$ the data exhibit a faster
rise towards the proton remnant direction than predicted by NLO QCD. 
This might be explained by a failure of the LO MC models to
describe the hadronisation corrections properly, by the inadequacy of
the photon PDF or by the need for higher order (HO) corrections.

The ZEUS Collaboration uses the measurement of jets in photoproduction
to extract values for $\alpha_S(E_T^{jet})$~\cite{bib:zeusgammapjets}. 
The jet cross section is fitted in each bin of $E_T^{jet}$ to

\begin{equation}
[{d\sigma}/{dE_T^{jet}}]_i
= A_i \alpha_S(E_T^{jet}) + B_i \alpha_S^2(E_T^{jet}),
\end{equation}

\noindent where the constants $A_i$ en $B_i$ are obtained from NLO QCD
calculations.  A fit of the energy-scale dependence of the extracted
$\alpha_S(E_T^{jet})$ to the renormalization group equation yields

\newpage

\begin{equation}
\alpha_S(M_Z) = 0.1224 \pm 0.0001 {\rm(stat.)} ^{+0.0022}_{-0.0019} {\rm
(syst.)} ^{+0.0054}_{-0.0042} {\rm (theo.)}.
\end{equation}

\noindent This is in agreement with the current world average of
$0.1183 \pm 0.0027$.
 
\section{Dijet electroproduction}

\begin{floatingfigure}{0.5\textwidth}
\begin{center}
\psfig{figure=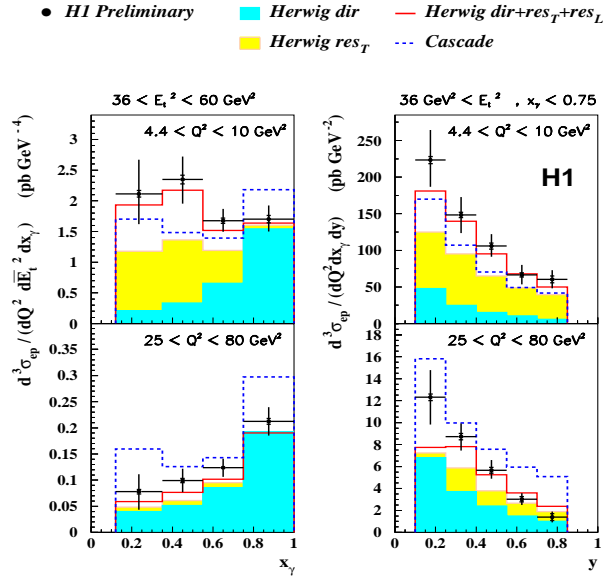,width=0.5\textwidth,height=7.7cm}
\end{center}

\caption{Differential cross section for dijet production as a
function of $x_\gamma$ or $y$ in bins of $Q^2$ and average $E_T^{jet}$.}

\label{fig:lowq2_dijets}
\end{floatingfigure}

When the virtuality of the exchanged photon is much larger than the
natural QCD mass scale ($Q^2 \gg \Lambda_{QCD}^2$), resolved photons
are in principle not needed.  However, LO QCD does not describe the
data on dijet production and HO effects therefore need to be included.

The H1 Collaboration compares two ap\-proach\-es~\cite{bib:h1dijets}. 
The first approach uses the HERWIG MC model~\cite{bib:herwig} based on
DGLAP QCD evolution with resolved photons.  Both transversely and
longitudinally polarized photons are needed to describe the measured
cross section, as shown in Fig.~\ref{fig:lowq2_dijets}.  The second
approach uses the CCFM evolution equations as implemented in the
CASCADE MC model~\cite{bib:cascade} with only direct photons.  This
model is equally well able to describe the data. It is therefore clear
that the ordering of a single DGLAP cascade needs to be broken, but
whether this can be achieved by introducing a second cascade
originating from the photon or whether new QCD dynamics is needed can
not be decided.

ZEUS computes the ratio of resolved to direct enhanced components of
the dijet cross section~\cite{bib:zeusdijets} defined as:

\begin{floatingfigure}{0.6\textwidth}
\begin{center}
\psfig{figure=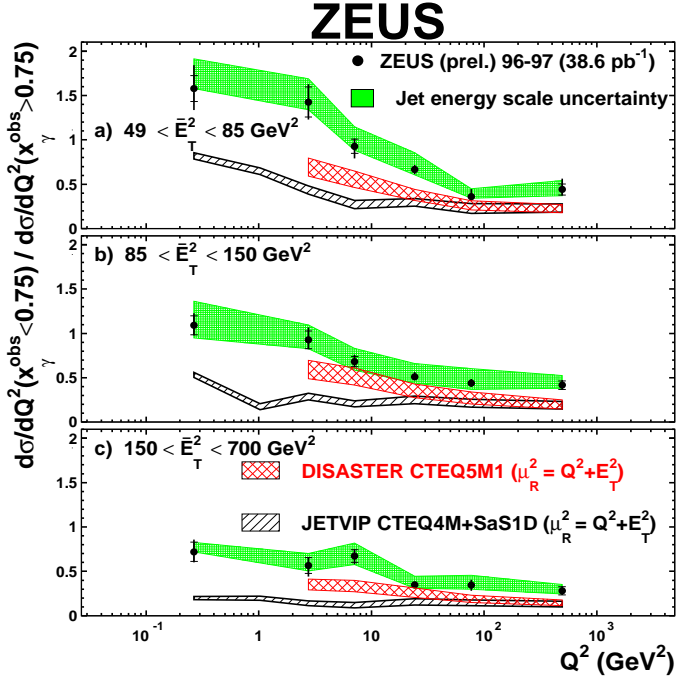,width=0.56\textwidth, height=9cm}
\end{center}

\caption{The ratio of resolved to direct enhanced components of the
dijet cross section as function of $Q^2$ in bins of average
$E_T^{jet}$.}

\label{fig:dis_dijets}
\end{floatingfigure}

\begin{equation}
R = \frac{\dfrac{d\sigma}{dQ^2}(x_\gamma^{jet} < 0.75)}
         {\dfrac{d\sigma}{dQ^2}(x_\gamma^{jet} > 0.75)}.
\end{equation}

\noindent Figure~\ref{fig:dis_dijets} shows $R$ as a function of $Q^2$
in bins of average $E_T^{jet}$.  Two NLO QCD models are used for
comparisons.  DISASTER++~\cite{bib:disaster++} only incorporates
pointlike photons, while JETVIP~\cite{bib:jetvip} implements both
direct and resolved photons.  Neither describes the ratio of resolved
to direct enhanced components well and although JETVIP should produce a
resolved enhanced component it actually performs worse than DISASTER++.
ZEUS also observes, however, that the direct enhanced component by itself is well
reproduced by both models.

\section{Inclusive jet electroproduction}

\begin{floatingfigure}{0.6\textwidth}
\begin{center}
\psfig{figure=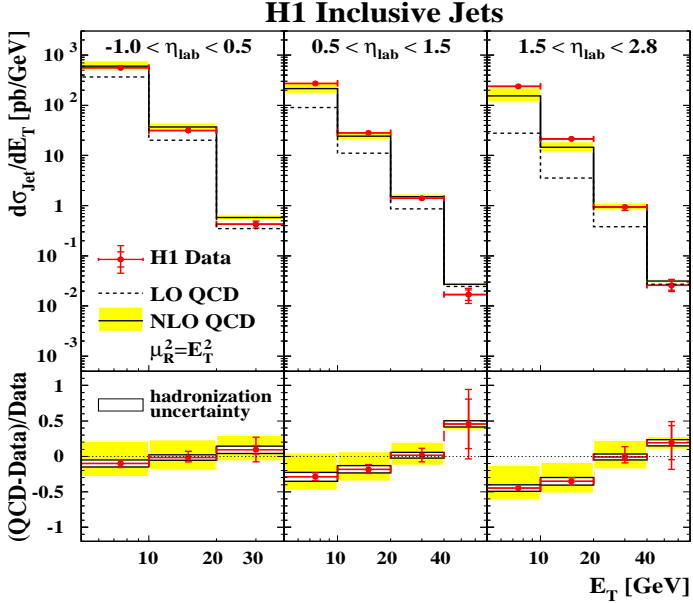,width=0.6\textwidth,height=8.2cm}
\end{center}

\caption{The inclusive jet cross section integrated over $5 < Q^2 <100
{\rm\ GeV}^2$ and $0.2 < y < 0.6$. The hatched band around the NLO
prediction represents the renormalization scale uncertainty.}

\label{fig:dis_jets}
\end{floatingfigure}

Inclusive jet cross sections obtained by H1~\cite{bib:h1disjets} are
shown in Fig.~\ref{fig:dis_jets}.  In the backward region and at all
$\eta_{lab}^{jet}$ for $E_T^{jet} > 20 {\rm\ GeV}$, NLO QCD
calculations by DISENT~\cite{bib:disent} describe the
${d\sigma^{jet}}/{dE_T^{jet}}$ distributions well.  Deviations are,
however, visible towards the proton remnant if both $E_T^{jet}$ and
$Q^2$ are small.  These discrepancies are accompanied by large
corrections between LO and NLO predictions.

Using a similar method as mentioned before, ZEUS has extracted values
for $\alpha_S(E_T^{jet})$~\cite{bib:zeusdisjets}.  The result is shown in
Fig.~\ref{fig:alphas} and is compared to the scale dependence from the
re\-normalization group equation using as input value the result obtained
from a fit to the measured ${d\sigma^{jet}}/{dQ^2}$ distribution
for $Q^2 > 500 {\rm\ GeV}^2$: 

\begin{equation}
\alpha_S(M_Z) = 0.1212 \pm 0.0017 {\rm(stat.)} ^{+0.0023}_{-0.0031} {\rm
(syst.)} ^{+0.0028}_{-0.0027} {\rm (theo.)}.
\end{equation}

\section{Conclusion}

\begin{floatingfigure}{0.5\textwidth}
\begin{center}
\psfig{figure=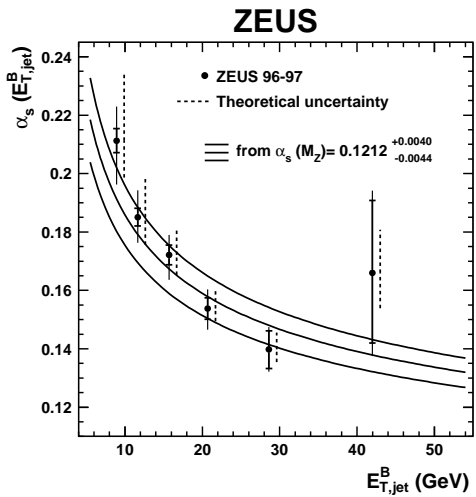,width=0.4\textwidth}
\end{center}

\caption{The $\alpha_S(E_T^{jet})$ values as determined from a QCD fit of
the measured jet cross section.}

\label{fig:alphas}
\end{floatingfigure}

The H1 and ZEUS Collaborations have measured jet production in $ep$
collisions with real and virtual photons in a large kinematic range.
Cross sections are obtained with high accuracy and fall over more than
six orders of magnitude as a function of transverse energy.

Competitive values for the strong coupling constant $\alpha_S$ are
obtained which are in agreement with the current world average.

NLO QCD calculations do a very good job in describing
jet cross section, with exceptions for forward jets at low $Q^2$ and
$E_T^{jet}$ and for the ratio of direct to resolved enhanced components in
dijet production.

This wealth of new high-precision data can, and should, be used in global fits of
photon and PDF's.

\section*{Acknowledgments}
I am indebted to all members of the H1 and ZEUS Collaborations who
contributed to these results by collecting and analysing the
experimental data.

\end{document}